\documentclass[preprint,showpacs,amsmath,amssymb]{revtex4}
\usepackage[dvips]{epsfig}
\usepackage{times,fancyhdr}
\usepackage{graphicx}
\usepackage{color}
\usepackage{hhline}

\def\e{\mathop{\rm \mbox{{\Large e}}}\nolimits}
\newcommand{\bn}[1]{\mbox{\boldmath $#1$}}
\newcommand{\noi}{\noindent}
\newcommand{\be}{\begin{equation}}
\newcommand{\ee}{\end{equation}}
\newcommand{\bc}{\begin{center}}
\newcommand{\ec}{\end{center}}
\newcommand{\bea}{\begin{eqnarray}}
\newcommand{\eea}{\end{eqnarray}}
\newcommand{\ba}{\begin{array}}
\newcommand{\ea}{\end{array}}

\def\1{_{1}}
\def\2{_{2}}

\begin{document}

\preprint{JAP(JR07-3622)}

\title{The Stationary Phase Method for a Wave Packet in a Semiconductor Layered System.
The applicability of the method.}

\author{H. Rodr\'{\i}guez-Coppola} \affiliation{Departamento de F\'{\i}sica Aplicada, Facultad de
F\'{\i}sica, Universidad de La Habana C.P. 10400, Cuba.}
\author{ L. Diago-Cisneros}\altaffiliation[{\bf Permanent}:]{Departamento de F\'{\i}sica Aplicada, Facultad de
F\'{\i}sica, Universidad de La Habana C.P. 10400, Cuba. } \affiliation{Departamento de F\'{\i}sica y
Matem\'aticas, Universidad Iberoamericana C.P. 01219, D.F. M\'exico}
\author{R. P\'erez-\'Alvarez}\altaffiliation[{\bf Permanent}:]{Departamento de F\'{\i}sica Te\'orica, Facultad de F\'{\i}sica,
Universidad de La Habana C.P. 10400, Cuba.}\affiliation{Facultad de Ciencias, Universidad Aut\'onoma del
Estado de Morelos C.P. 62209, Cuernavaca, M\'exico.}

\hyphenation{par-ti-ci-par he-te-roes-truc-tu-ra des-cri-bir com-pa-ra-ti-va-men-te re-fe-ren-cia
ex-pe-ri-men-tal pro-ble-ma}

\begin{abstract}
Using the formal analysis made by Bohm in his book, {\em ``Quantum theory"}, Dover Publications Inc. New
York (1979), to calculate approximately the phase time for a transmitted and the reflected wave packets
through a potential barrier, we calculate the phase time for a semiconductor system formed by different
mesoscopic layers. The transmitted and the reflected wave packets are analyzed and the applicability of
this procedure, based on the stationary phase of a wave packet, is considered in different conditions.
For the applicability of the stationary phase method an expression is obtained in the case of the
transmitted wave depending only on the derivatives of the phase, up to third order. This condition
indicates whether the parameters of the system allow to define the wave packet by its leading term. The
case of a multiple barrier systems is shown as an illustration of the results. This formalism includes
the use of the Transfer Matrix to describe the central stratum, whether it is formed by one layer (the
single barrier case), or two barriers and an inner well (the DBRT system), but one can assume that this
stratum can be comprise of any number or any kind of semiconductor layers.
\end{abstract}

\maketitle

\section{INTRODUCTION}
\noi In the last century the calculation of the time spent by a particle when passing through a
potential barrier was, for a long time, one of the basic and controversial problems since the early days
of Quantum Mechanics. When the issue of the delay time of a transmitted wave packet through a potential
barrier was under investigation by MacColl\cite{MColl} and later by Hartman,\cite{Hartman} using the
Wigner's phase time introduced in nuclear physics, the striking superluminal effect arose immediately.
At this time the question ``how much time does tunnelling take" was loosely formulated.\cite{MColl}
Early answers to this problem \cite{Bohm,Wigner55} and alternative proposals
\cite{Bohm,Wigner55,Smith,Baz,Buttiker,Hauge89,Winful} run from pure semiclassical to fully quantum
mechanical models. Nowadays the impressive number of low-dimensional semiconductors devices brought a
new urgency to the essential measurement and/or modelling of tunnelling time for charge carriers motion.
The last can be seen reflected in the large presence of publications. \\ \noi  Early in the '$90$s, real
experiments on photon-twins interference and on optical pulses propagation\cite{Steinberg93,Spielmann94}
measured the delay time, in a simple and direct way, at first. On the other hand, most of the available
experimental setups, pretending to be relevant to the tunnelling issue, actually involve other times
derived from scape and/or decay phenomena. In this sense, their results are not able to identify real
tunnelling time scale, and consequently should be questionable as potentially misleading. Authentically
connected to the tunnelling process delay measurements,\cite{Steinberg93,Spielmann94,Nimtz02} and
uncommonly good agreement with some of them,\cite{Steinberg93,Spielmann94} found within the phase-time
model,\cite{PPP3} are striking developments from the days of the lively debate on these matters appeared
during the late $1980$'s. Approximate\cite{Esposito01} and multiband \cite{LDC1} phase time calculations
in different systems, confirm experimental results reported in
Ref.[\onlinecite{Steinberg93,Spielmann94,Nimtz02}] and in Ref.[\onlinecite{Heberle94}], respectively.
The robustness of the phase time approximation was assured by its consistency with the Maxwell's
equations predictions.\cite{PPP-Herbert07} Being largely stimulated by the success of the phase-time
conception, we had applied here the Stationary Phase Method (SPM),\cite{SPM} firstly to evaluate the
phase time as it straightforwardly deals with both initial (incident) and final (transmitted and
reflected) dispersion phase amplitudes and finally to further study this magnitude to know its
applicability more closely. \\ \noi In the last years some other authors have studied these problems in
many respects.\cite{NgChan, Yamada,PPP1,PPP2} At the same time, in Ref.[\onlinecite{NgChan}] several
references of different applications of the tunnelling process in different low dimensional
semiconductor devices are given. This is the reason to further study this magnitude to know its
applicability more closely.  \\ \noi In the present paper the formal analysis of Bohm\cite{Bohm} is used
to determine the transmitted wave packet and, properly, the phase time in an arbitrary semiconductor
layered system for a charge carrier trespassing the structure. The point is that Bohm got the leading
term of the transmitted wave packet and from it, got an expression to calculate the phase time, and
these approximations not always are good. To reach when they are good and when do not is our task. The
system could be described by a single band or multiband models. In figure \ref{fig:A} the general view
of the system under study is depicted, no matter how many layers are included in the group $M$. \\ \noi
Using the Transfer Matrix (TM) method,\cite{RPA1} the wave packet reflected and transmitted are
obtained, using the SPM\cite{SPM} to solving the integrals for these waves. This method lead us to an
applicability condition for it and, properly, the phase time as a function of the parameters of the
system. The application of the resulting expressions for the Schr\"odinger single band case is given and
some results obtained for the double barrier resonant tunnelling semiconductor structure (DBRT) are
given for illustration. Some comments were included to extend these results to the case of a system
described by $N$ second order differential system.

\section{THE FORMAL ANALYSIS}
\begin{figure}[!h]
\hspace*{1.2in}\includegraphics[width=3.3in,height=2.5in]{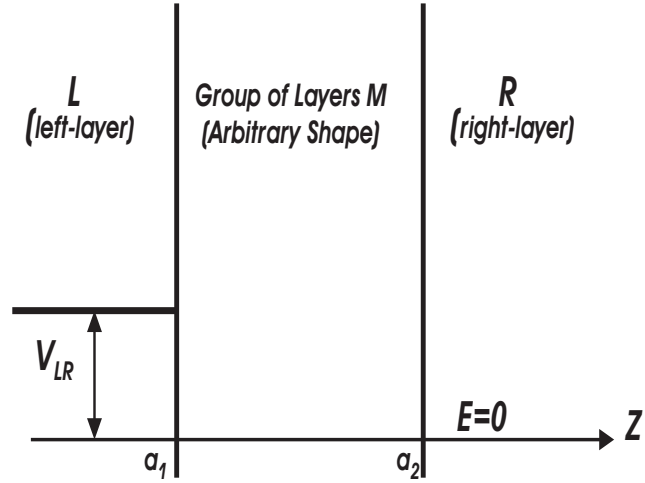}\caption{General view of the Potential
System under study. Layers $L$ and $R$ must not be equal necessarily. $V_{LR}$ is the potential difference
between the $L$ band edge and the and the $R$ band edge taken as energy reference level. In $z$-axis the
interfaces are named as $a_1$ and $a_2$.} \label{fig:A}
\end{figure}

\noi In the system depicted in figure \ref{fig:A} the Schr\"odinger wavefunction can be written as:
\begin{eqnarray}
\label{SchrA1}
\psi_o(z,p_R,t) & = & \e^{-iE(p_R)t/\hbar}\phi_o(z,p_R) \\
\label{SchrA} \phi_o(z,p_R) & = & \left\{\begin{array}{ll} D(p_R)\left|\left|\begin{array}{l} 1 \\
i\left(\sqrt{p_R^2 - 2m^*V_{LR}}/\hbar\right) \\ \end{array}\right|\right|\e^{iz\sqrt{p_R^2 -
2m^*V_{LR}}/\hbar} +  \\ \hspace*{5mm}+ F(p_R)\left|\left|\begin{array}{l} 1 \\ -i\left(\sqrt{p_R^2 - 2m^*V_{LR}}/\hbar\right) \\
\end{array}\right|\right|\e^{-iz\sqrt{p_R^2 - 2m^*V_{LR}}/\hbar}
& \mbox{$z < a_1$} \\  &  \\ \bn{M}(z,a_1-)\bn{\psi}(a_1-,p_R) & \mbox{$a_1 < z < a_2$} \\  & \\
A(p_R)\left|\left|\begin{array}{l} 1 \\ i(p_R/\hbar) \\
\end{array}\right|\right|\e^{ip_Rz/\hbar} & \mbox{$z > a_2$} \end{array}\right.;
\end{eqnarray}
\noi where the TM of wavefunction and derivative\cite{RR} was included to describe the central layer
$M$.\cite{RPA1} As this part can be arbitrary, the expression for the TM will depend on the form of the
potential of this layer. Also it was written $p_R =
\sqrt{2m^*E}$. The wavefunction for the layer $L$ was also written in terms of $p_R$ for convenience. In
doing this, the potential $V_{LR}$ was defined (see caption of Figure \ref{fig:A}). \\
\noi The wave packet is obtained, for different values of coordinate $z$ by forming the expression:
\be\label{Schr2}\Psi(z,t) = \int_{-\infty}^{\infty}dp_Rf(p_R - p_{Ro})\psi_o(z,p_R,t),\ee \noi where
function $f(p_R - p_{Ro})$ is a shape function which peaks at the value $p_{Ro}$ and rapidly goes
to zero for large values of the difference $p_R - p_{Ro}$, then the integral limits can be extended to $\pm\infty$. \\
\noi In the case of the transmitted wave, taking $D(p_R) \equiv 1$ as a condition of normalization of the
wavefunction used to form the wave packet, one obtains for $\Psi_R(z,t)$ for $z$ in region $R$:
\be\label{Schr3}\Psi_R(z,t) = \int_{-\infty}^{\infty}dp_Rf(p_R - p_{Ro})B(p_R)\e^{\{i(zp_R/\hbar) +
i\phi_t(p_R) -i\left(E(p_R)\right)t/\hbar\}}.\ee \noi The normalization condition means that the incident
wave to form the packet is normalized in the $L$ region. Here it was written $A(p_R) \equiv
B(p_R)\e^{i\phi_t(p_R)}$ using $\phi_t(p_R)$ as the phase of the transmitted wave amplitude. For the case
$N \geq 2$ ({\em i. e.} a physical system described by two or more coupled differential equations), the
condition of normalization have to be released because it is necessary to write the spinor as a part of
the wave function.\cite{NN} In parameter $A(p_R)$ the matching of the different layers in the structure
is included. For $N \geq 2$ this parameter is a vector, then this matching process appears in the
coefficient $B(p_R)$ and in the phase $\phi_t(p_R)$ which in the multiband case must be calculated by
components and no matrix expression can be given. \\ \noi Considering the SPM to perform the integral in
the case $N = 1$ one has to expand in Taylor's series the exponent in (\ref{Schr3}) (which we called
$\Theta_T(p_R)$) and taking the value of $p_R$ which produces an extreme for the exponent ($p_{Ro}$) as
the approximation, one obtains:
\begin{eqnarray}
\label{Schr3A}  \Theta_T(p_R) & = & i(zp_R/\hbar) + i\phi_t(p_R) -i\left(E(p_R)\right)t/\hbar. \\
\mbox{} &  & \mbox{\hspace*{8mm}The definition of the exponent} \nonumber \\
\label{Schr3B} \Theta_T(p_{Ro}) & \approx & i(zp_{Ro}/\hbar) -i\left(E(p_{Ro})\right)t/\hbar +
i\phi_t(p_{Ro}) + O(p^2_{Ro}) + \cdots \\
\mbox{} &  & \mbox{\hspace*{8mm}The Taylor series up to second order}, \nonumber  \\
\label{Schr4} \Psi_R(z,t) & = & G(p_{Ro})\e^{i\left[-(tE(p_{Ro})/\hbar) '+
(zp_{Ro}/\hbar)\right]}\e^{i\left(\phi_t(p_{Ro}) + \pi/4\right)} \\
\mbox{} &  & \mbox{\hspace*{8mm}The leading term of the wave packet}, \nonumber \\
\label{Schr5} G(p_{Ro}) & = & \frac{\sqrt{2\pi}B(p_{Ro})}{\left[\left(d^2\phi_t(p_R)/dp_R^2\right)_{p_{Ro}}
- \left[1/(m^*v_{gR})\right]\left(d\phi_t(p_R)/dp_R\right)_{p_{Ro}}\right]} \\
\mbox{} &  & \mbox{\hspace*{8mm}The coefficient of the leading term of the wavepacket}. \nonumber
\end{eqnarray}
\noi Here we have considered that $f(p_{Ro} - p_{Ro}) = 1$ and we use $v_{gR}$ as the group velocity of the
packet in layer $R$. Expression (\ref{Schr4}) is the leading term of the transmitted wavefunction, obtained
by making this approximation. The coefficient of this wavefunction is given by (\ref{Schr5}) written in
terms of the phase delay of the transmitted wave. \\ \noi The applicability of the SPM takes into account
that it uses the Taylor expand of the exponent and neglects the terms from the second order. This leads to
write:
\begin{eqnarray}
\label{phase1}
\Theta_T(p_R) & = & \Theta_T(p_{Ro}) + \Theta_T'(p_{Ro})(p_R - p_{Ro}) +
\frac{1}{2}\Theta_T''(p_{Ro})(p_R - p_{Ro})^2 + \cdots. \\
\label{phase2} \Theta_T(p_R) & = & \Theta_T(p_{Ro}) + \frac{1}{2}\Theta_T''(p_{Ro})(p_R - p_{Ro})^2 +
\cdots; \mbox{\hspace*{8mm} as is an extreme.} \\
\label{phase4}\frac{\Theta_T'''(p_{Ro})(p_R - p_{Ro})^3}{\Theta_T''(p_{Ro})(p_R - p_{Ro})^2} & = &
\frac{\Theta_T'''(p_{Ro})(p_R - p_{Ro})}{\Theta_T''(p_{Ro})} < \epsilon;  \mbox{\hspace*{8mm}with
$\epsilon\rightarrow 0$.}
\end{eqnarray}
\noi Evaluating the derivatives of the exponential phase (\ref{phase4}) in terms of the derivatives of
the phase of the transmitted wave one obtains as the condition for the applicability of the SPM the
expression: \be\label{Schr6}\Sigma(p_R) = \frac{\hat{\Omega}_n(p_R)}{\hat{\Omega}_d(p_R)} =
\large{\frac{\left|\left(\frac{d^3\phi_t(p_R)}{dp_R^3}\right)_{p_R=p_{Ro}}\right|}{\left|-
\frac{1}{p_{Ro}}\left(\frac{d\phi_t(p_R)}{dp_R}\right)_{p_R=p_{Ro}} +
\left(\frac{d^2\phi_t(p_R)}{dp_R^2}\right)_{p_R=p_{Ro}}\right|^{3/2}} \ll 1}.\ee \noi This is the main
contribution of this paper. This condition evaluates the applicability of the SPM and points over the use
of the phase delay time for every group of values of the parameters of the system.\\ \noi Nevertheless,
this expression (\ref{Schr6}) has the numerator and the denominator dimensional and the quotient non
dimensional, then to properly compare these expressions it is better to multiply by $p_{Ro}^3$ both,
numerator and denominator. This lead us to:
\begin{eqnarray}
\label{SS1} \Sigma(p_R) & = & \frac{\hat{\Omega}_n(p_R)\cdot p_{Ro}^3}{\hat{\Omega}_d(p_R)\cdot p_{Ro}^3}
= \frac{\Omega_N(p_R)}{\Omega_D(p_R)} \\ \label{SS2}\Omega_N(p_{Ro}) & = &
\left|p_{Ro}^3\left(\frac{d^3\phi_t(p_R)}{dp_R^3}\right)_{p_R=p_{Ro}}\right| \\
\label{SS3}\Omega_D(p_{Ro}) & = & \left|- p_{Ro}\left(\frac{d\phi_t(p_R)}{dp_R}\right)_{p_R=p_{Ro}} +
p_{Ro}^2 \left(\frac{d^2\phi_t(p_R)}{dp_R^2}\right)_{p_R=p_{Ro}}\right|^{3/2}
\end{eqnarray}
\noi The phase time for the transmitted wave is obtained from the condition of stationary phase of the
exponential in the integral (\ref{Schr3}). After including the matching at layer boundaries, one has for
the phase of the exponential in (\ref{Schr4}) the expression:
\begin{eqnarray}
\label{Schr6.6}\alpha_T(p_R,t) & = & \frac{zp_R}{\hbar} + \phi_t(p_R) - \frac{t}{\hbar}E(p_R), \\
\label{Schr7} \frac{d\alpha_T}{dp_R} = 0 & \Longrightarrow & \tau_T =
\frac{m^*\hbar}{p_{Ro}}\left(\frac{d\phi_t(p_R)}{dp_R}\right)_{p_R=p_{Ro}};
\end{eqnarray}
\noi which is the formula to evaluate the phase delay time of the transmitted wave.\cite{Bohm} In
(\ref{Schr6.6}) was taken, as Bohm did in his book, $z \equiv \Delta z = 0$ because it refers to the phase
between group $M$ of layers and layer $R$, {\it i. e.}, the wave packet reaches the same position, later
than if there were no dispersion potential causing the wave to be reflected. In this sense, the phase of
layer $R$ differentiates, bearing a term that comprises wave packet's evolution delay information. \ \noi
For the case of $N \geq 2$ bands, the whole analysis cannot be generalized for the present scheme from the
case $N = 1$ because the step of converting a complex number from $a + ib$ to $\rho\e^{i\phi}$ cannot be
performed in matrix notation and one must passes to $2N\times2N$ components. Further investigation is
required to write close expressions in this case. This is important because there are several problems
described by the standard Sturm-Liuoville $N\times N$ differential equation system\cite{RPA1} of great
practical interest. Models as that due to Bogoliubov for superconductor excitations description\cite{Bog}
could be treated as well.
\\ \noi A simple consideration of closeness between the phase-time model and the
dwell time (within its phase-time probabilistic average formulation \cite{Hauge96,Diosdado05}), dispose us
to speculate that the requirement (\ref{Schr6}) should be readily suited to it, with minor changes. We are
interested in comparing these two possible conditions to get light into the use of different times for
tunnelling processes.

\section{RESULTS AND DISCUSSION}
\begin{figure}[!h]
\hspace*{1.2in}\includegraphics[width=3.3in,height=2.5in]{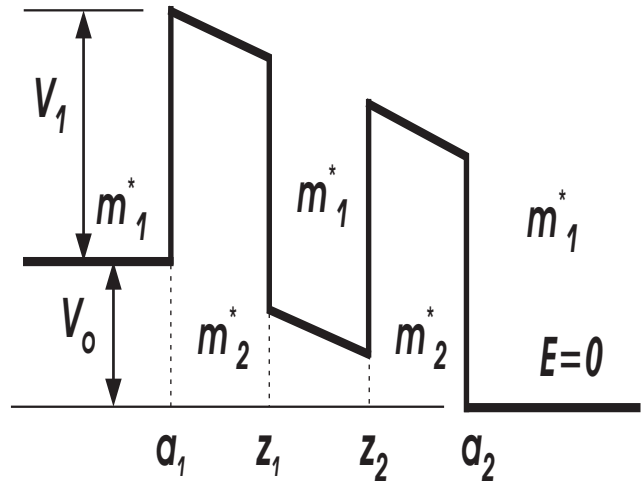}\caption{General view of the DBRT system
under study. Here $V_o$ is called the $V_{LR}$ of Figure \ref{fig:A}. Points $a_1$ and $a_2$ are here the
same as in Figure \ref{fig:A}, then points $z_1$ and $z_2$ and potential $V_1$ belong to layer $M$ of
Figure\ref{fig:A}.} \label{fig:B}
\end{figure}

\begin{figure}[!h]
\hspace*{1.2in}\includegraphics[width=3.3in,height=3in]{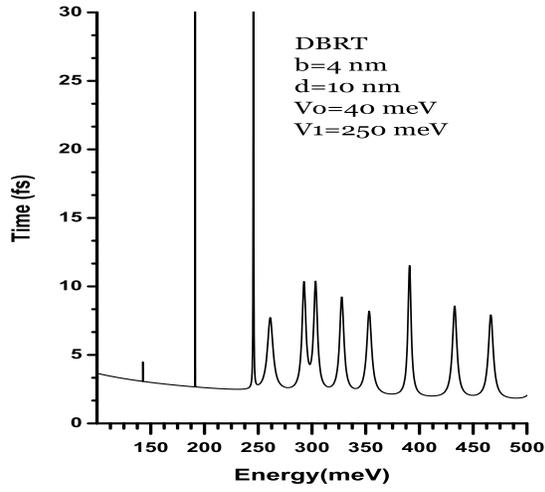}\caption{Phase delay time calculated by
(\ref{Schr7}) for the DBRT with the parameters given in Table \ref{T1}. For $E < V_o$ the picks correspond
to energies of the inner well.} \label{fig:C}
\end{figure}

\noi The application of this formal analysis to different physical systems allows one to determine
whether the phase time can be applied to a given system and to obtain it from the wavefunction. As an
illustration we applied this procedure to the case of a double barrier resonant structure device in
$GaAs/AlGaAs$ considering the parameters shown in Table \ref{T1}. \\ \noi The potential of the system is
depicted in Figure \ref{fig:B}, where the extreme left and right layers were considered as metallized
contacts, which are
semiconductors ($GaAs$) with flat band and an electric field applied to the structure. \\
\noi Using (\ref{Schr7}), after making the matching considering the differences of masses in each layer
by using the TM algorithm,\cite{RPA1} the phase delay time has the behavior depicted in figure
\ref{fig:C} as a function of the energy of the incident wave. \\ \noi Our results for the phase time
depicted in Figure \ref{fig:C} are of the same order of magnitude of other calculations and the behavior
of the phase time is as others achieved, as can be seen in table \ref{TabTT} for electrons and photons
in similar system, reported elsewhere.\cite{PPP3,Yama98,Porto94,Longhi02} Several methods were used by
these authors, namely: lifetime,\cite{Yama98} dwell time,\cite{Yama98} Wentzel-Kramer- Brillouin (WKB)
quasi-classical approximation\cite{Porto94} and phase time.\cite{PPP3} In the case of photons, the
reported values correspond to $1.5\;\mu$m optical pulse wavelength, propagating through double-barrier
photonic band gap (FBG).\cite{Longhi02} In this table are included some useful data as if there is
applied electric field, if the results were achieved theoretical or experimentally and the model used to
perform the calculation.
\\ \noi The application of the SPM to this system is governed by expression (\ref{Schr6}) and in Figures
\ref{fig:D} a, b, c, d and e are shown separately the numerator $\Omega_N(p_R)/10^6$, the denominator
$\Omega_D(p_R)/10^{14}$ and the quotient $\Sigma(p_R)$ of the applicability condition (\ref{SS1}) for
different energy ranges. This is the main result of this paper, because the phase delay time
approximation is already known, but (\ref{SS1}) is not used to assure its application to different
systems. It is easily seen that in all graphics $\Omega_N$ is under $\Omega_D$, so the procedure and the
phase delay time are valid for all the energy range of interest. Nevertheless, there is an isolated
point,
seen in figure 4e) that goes over unity, and makes the SPM and the phase delay time inapplicable. \\
\noi This analysis allows to say that this definition of time is good enough for many useful analysis at
all energy ranges.\\ \noi As a conclusion we have calculated the phase delay time in a system of
semiconductor layers, illustrating with the simple case of a DBRT system described by the Schr\"odinger
equation with an electric field applied and some light about the applicability of this definition of time
is given by considering the condition obtained for the use of the SPM in reaching the transmitted wave
packet. It is clear that one has to apply the applicability condition in each case under study to assure
that the phase delay time is good in the conditions of each concrete problem. It is also an interesting
guess to extrapolate the applicability condition obtained (\ref{Schr6}) for the phase time, to the case of
the dwell time in its probabilistic average formulation\cite{Hauge96,Diosdado05} with minor changes. \\
\noi Also in this paper some considerations were made to extend these formulae to the case of systems with
$N$ second order coupled differential equations in which some of the algebra must be done in matrix
notation and other cannot. This application leads to individual results for each component separately and
after that one can rebuild the matrices. Expression (\ref{Schr6}) is valid for each one of the components
and must be obtained and evaluated individually. The application of these results to the case of $N$ second
order differential equations is in progress.

\newpage

\begin{table}[!ht]
\caption{Parameters of the DBRT considered in the calculation sketched in Figure \ref{fig:B}}\label{T1}
\begin{center}
\begin{tabular}{|clc|}\hline\hline
No & Parameter & Value \\ \hline 1 & Barrier Height ($V_1$) & 250 meV \\ 2 & Difference between the band
edges & \mbox{} \\ \mbox{} & of sides $L$ and $R$ ($V_o$) & 40 meV \\ 3 & Barrier width $b = z_1 - a_1$ & 40 \AA \\
4 & Well Width $d = z_2 - z_1$ & 100 \AA \\ 5 & $m_1^*$ in units of $m_o$ & 0.066 \\ 6 & $m_2^*$ in
units of $m_o$ & 0.8 \\ \hline
\end{tabular}
\end{center}
\end{table}

\newpage

\begin{table}[!ht]
\caption{\label{TabTT}Time scale for tunneling}
\begin{center}
\begin{tabular}{|lclccclc|}\hline\hline
System & Potential & Structure & Data Source & Resonance &  Bias & Time & Value of Time  \\
\mbox{} & \mbox{} & \mbox{} & \mbox{} & [eV] & [meV] & \mbox{} & [ps] \\\hline
electrons & DBRT & Al$_{0.3}$Ga$_{0.7}$As/GaAs & Teo. Ref[\onlinecite{Yama98}] & $0.05$ & - & life & $5.7$ \\
electrons & DBRT & Al$_{0.3}$Ga$_{0.7}$As/GaAs & Teo. Ref[\onlinecite{Yama98}] & $0.18$ & - &  life & $2.9$ \\
electrons & DBRT & Al$_{0.3}$Ga$_{0.7}$As/GaAs & Teo. Ref[\onlinecite{Yama98}] & $0.05$ & - &  dwell & $5.7$ \\
electrons & DBRT & Al$_{0.3}$Ga$_{0.7}$As/As & Teo. Ref[\onlinecite{Yama98}] & $0.18$ & - &  dwell & $2.9$ \\
electrons & DBRT & Ga$_{0.47}$In$_{0.53}$As/Al$_{0.48}$In$_{0.52}$As & Teo. Ref[\onlinecite{Porto94}]
          & - & $40$ & WKB & $0.5$ \\
electrons & DBRT & Al$_{0.3}$Ga$_{0.7}$As/GaAs & Teo. Ref[\onlinecite{PPP3}] & $0.11$ & - & phase &
          $0.02$ \\
electrons & DBRT & Al$_{0.3}$Ga$_{0.7}$As/GaAs & Teo. Ref[\onlinecite{PPP3}] & - & - & phase &
          $\approx 0.02$ \\
photons & FBG & mono-mode optical fiber & Exp. Ref[\onlinecite{Longhi02}] & - & - & traversal &
          $180$ \\
photons & FBG & mono-mode optical fiber & Teo. Ref[\onlinecite{Longhi02}] & - & - & phase &
          $300$ \\ \hline
\end{tabular}
\end{center}
\end{table}

\newpage

{\bf Figure Captions}

\begin{figure}[!h]
\hspace*{1.2in}\includegraphics[width=3.3in,height=3in]{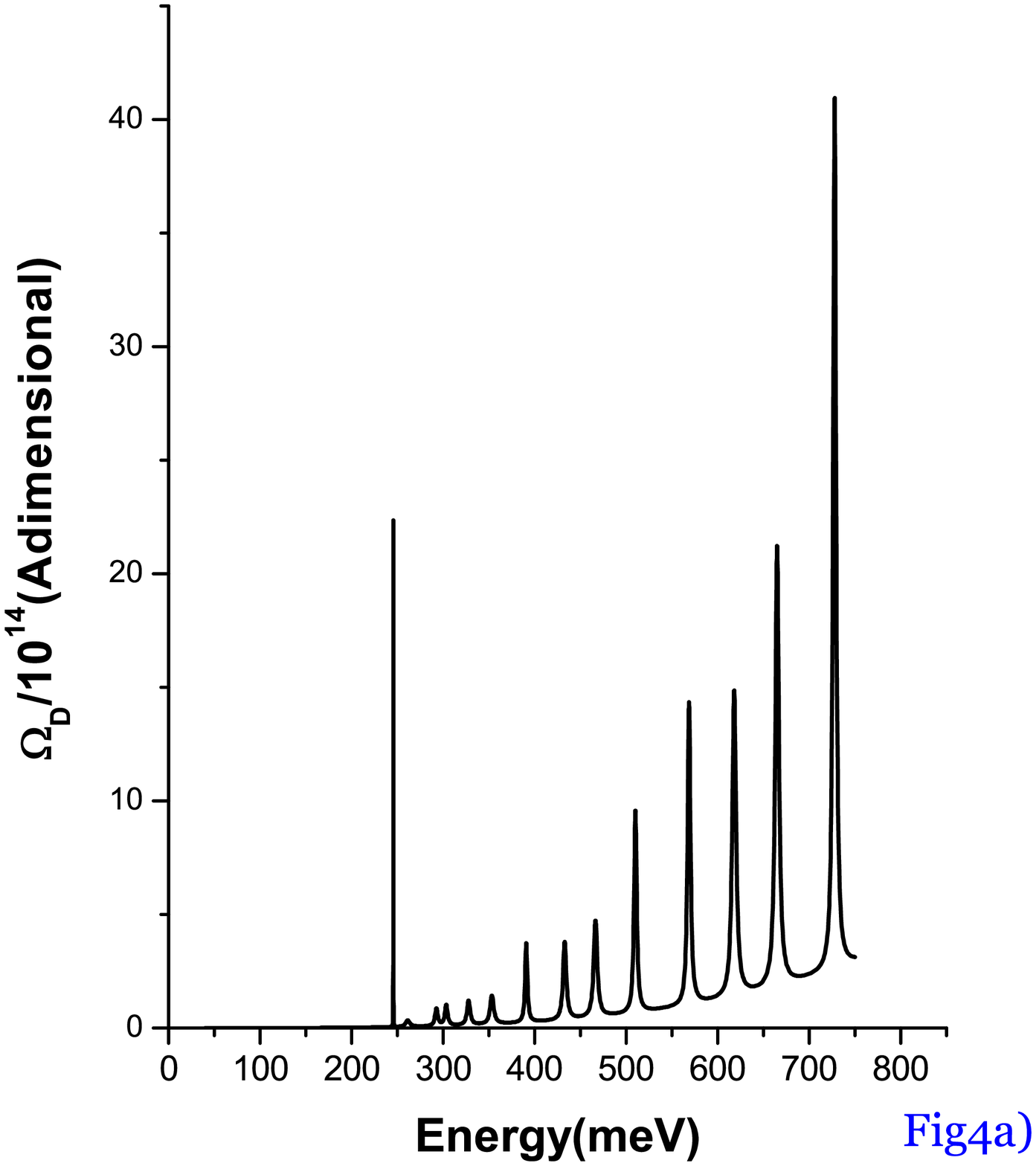}
\end{figure}
\begin{figure}[!h]
\hspace*{1.2in}\includegraphics[width=3.3in,height=3in]{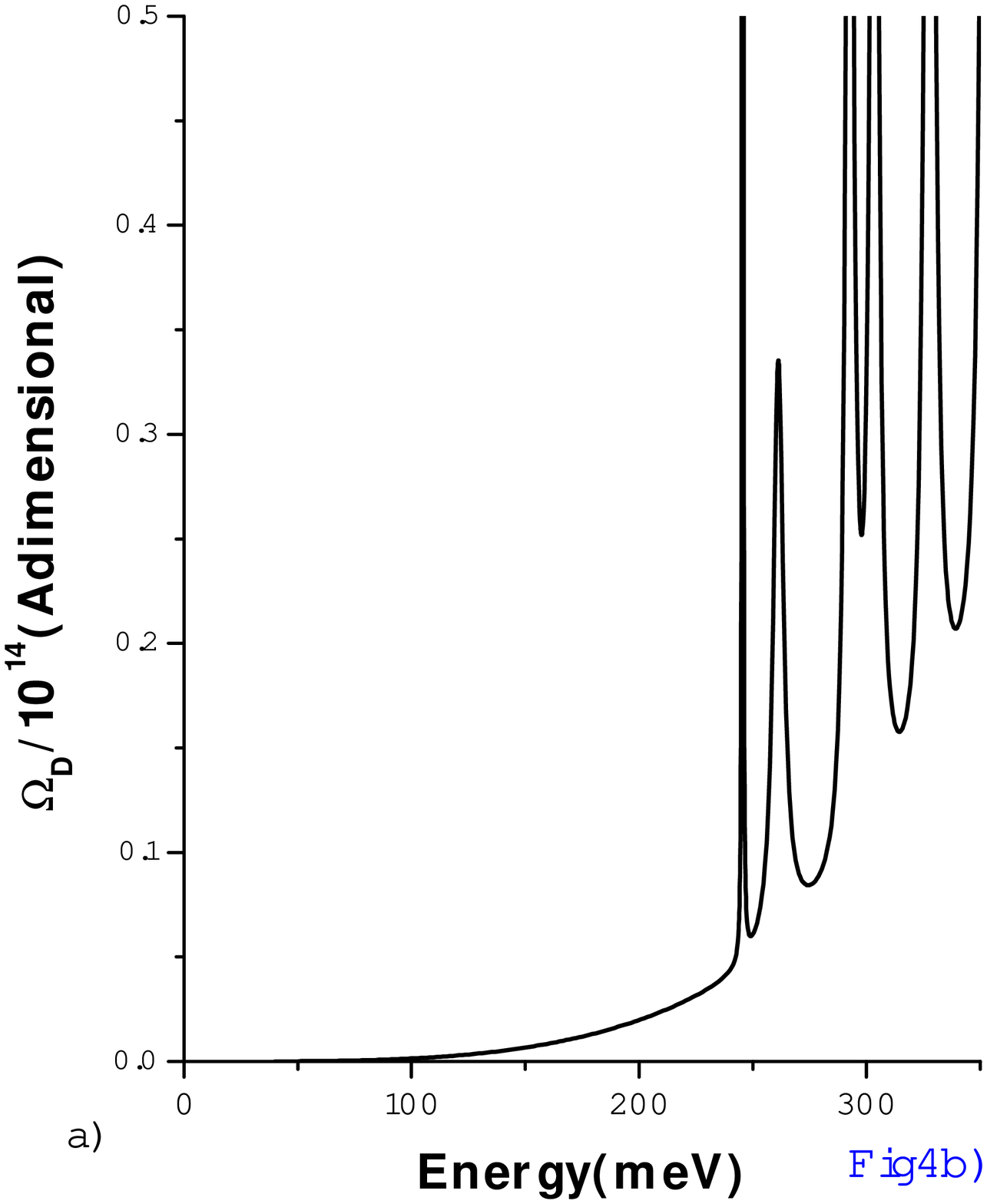}
\end{figure}
\begin{figure}[!h]
\hspace*{1.2in}\includegraphics[width=3.3in,height=3in]{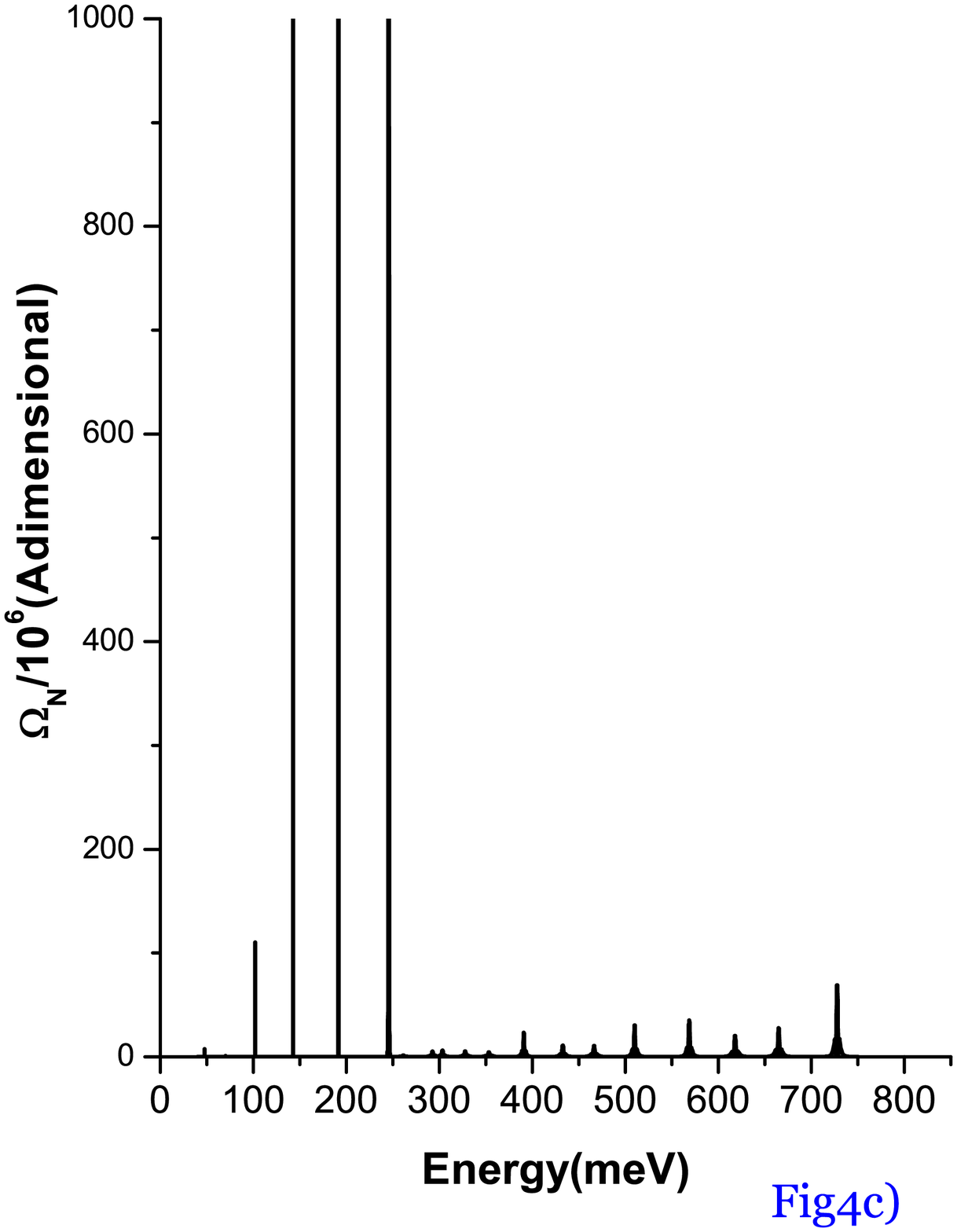}
\end{figure}
\begin{figure}[!h]
\hspace*{1.2in}\includegraphics[width=3.3in,height=3in]{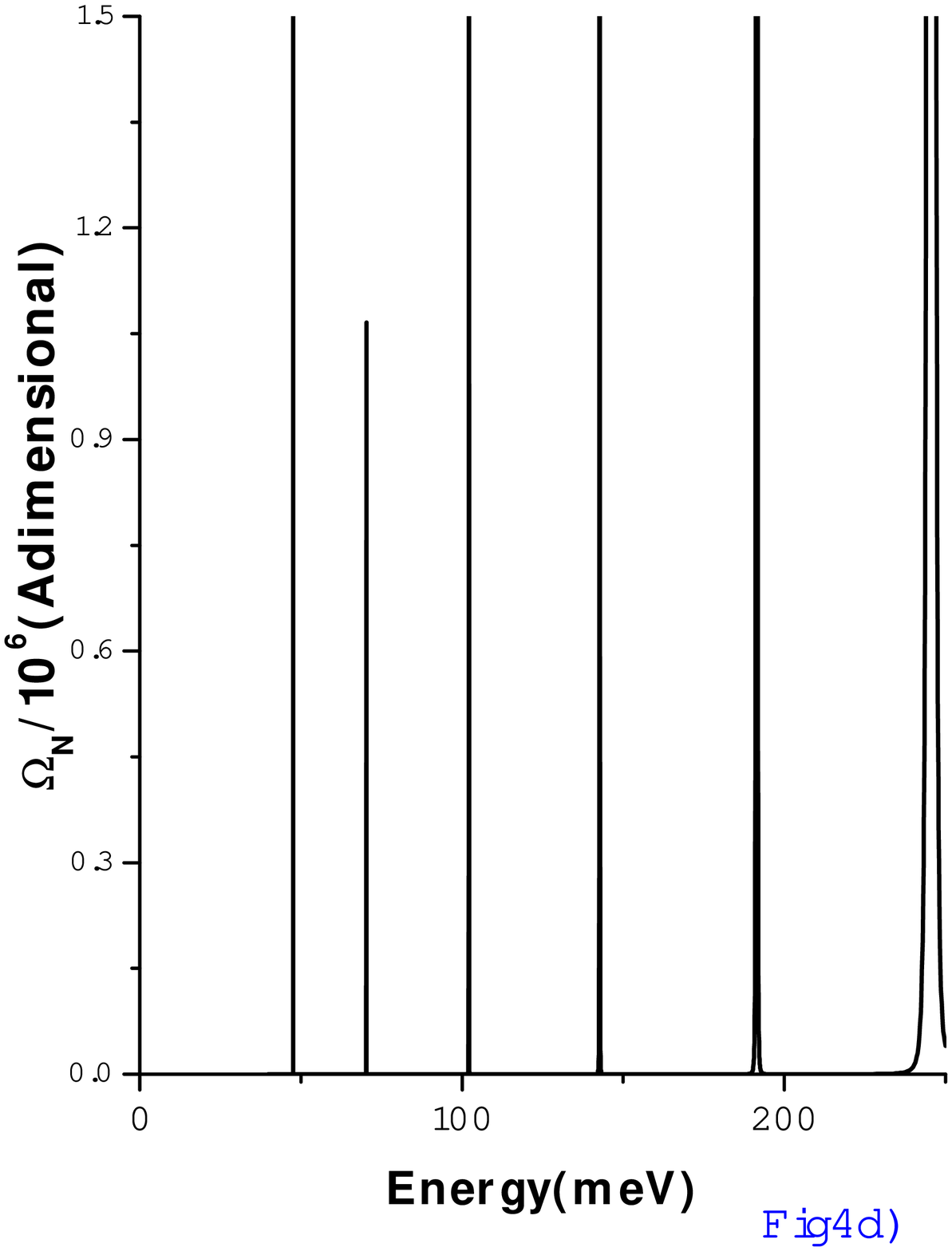}
\end{figure}
\begin{figure}[!h]
\hspace*{1.2in}\includegraphics[width=3.3in,height=3in]{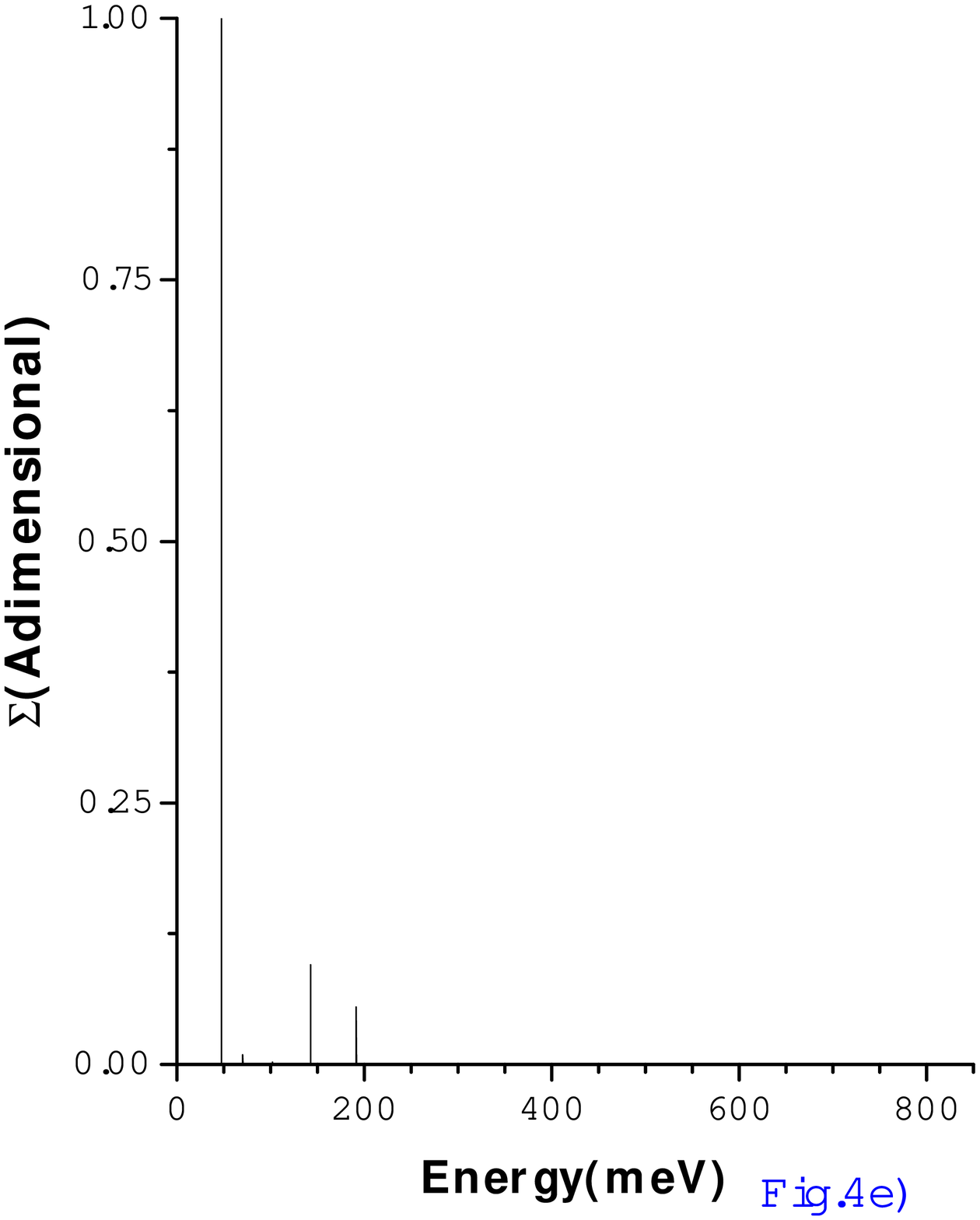}\caption{Evaluation of the condition of
applicability of the SPM (\ref{Schr6})in the energy range. The shown graphs are: a) Denominator
($\Omega_D(p_R)/10^{14}$) versus energy, b) The same as a) but in a particular energy range. c) Numerator
($\Omega_N(p_R)/10^6$) vs energy, d) The same as c) but in a smaller energy range. and e) Quotient
($\Sigma(p_R)$) versus energy in the whole energy range.} \label{fig:D}
\end{figure}


\begin{thebibliography}{40}
\bibitem{RR} There are several definitions of TM. Normally we use three of them. One transferring the
wavefunction and the derivative in a domain; other transferring the wavefunction and the linear form
which must be continuous at the interfaces and finally one transferring the coefficients of the
wavefunction in the representation of propagating modes. (See \onlinecite{RPA1}). In the present paper
we use the first TM mentioned.
\bibitem{NN} This problem is not trivial. Usually the multiband cases are taken by considering all
components but one zero, which is a particular case. Our analysis (see Ref.[\onlinecite{LDC1}]) usually
releases this consideration and all components are non zero. One must put the coefficient of the
incident wave in layer $L$ as normalized only. This means that our incident wave is a combination of all
components.
\end{thebibliography}
\end{document}